\documentclass[12pt]{EJLateX/ejtex}
\usepackage{amsmath}
\EJissue{6}
\EJdate{Autumn 2013}

\begin{document}

\begin{EJarticle}[High Energy Quantum Mechanics]{The High Energy Interpretation of Quantum Mechanics}{Stefano Forte}{Dipartimento di Fisica, Universit\`a di Milano and
INFN, Sezione di Milano,\\ Via Celoria 16, I-20133 Milano, Italy\\ stefano.forte@mi.infn.it}{Dipartimento di Fisica, Universit\`a di Milano and
INFN, Sezione di Milano} 

\begin{EJabstract}
We address the issue of interpretation of quantum mechanics by
asking why the issue never arises in the description of high-energy
interactions. We argue that several tenets of quantum mechanics,
specifically the collapse of the wave function, follow directly once
accepts the essential randomness of fundamental interaction events. We
then show that scale separation of fundamental interactions ensures that the
measurement can be unambiguously separated from the quantum
events. Finally, we argue that the fundamental symmetries of space and
time guarantee the existence of a unique preferred basis. We argue
that this set of ideas might lead to an interpretation of quantum
mechanics, or rather, show in which sense an ``interpretation'' is 
necessary. 
\end{EJabstract}

\section{The Quantum World and its ``Paradoxes'' }
It is a commonplace statement that the laws of quantum mechanics are
strange and to some extent paradoxical: a statement which is 
reflected not only in the titles of 
popular science books (see e.g.~\cite{arroyo,zeilinger}) but also in
chapter headings of textbooks (e.g. ``Abandoning realism'' in Chapter
3.7 of~\cite{weinbergqm}) or even papers
(see~\cite{laloe}
on quantum
mechanics. Of course, in popular science,  relativity, say, 
(even special relativity)
is also often described as puzzling and paradoxical. However, as any physicist
(in fact, any decent physics undergraduate) knows, special
relativity expresses the kinematic properties of physical laws in a
way which is neither more complicated nor less intuitive than
Galilean relativity. Indeed, both special relativity and Galilean
relativity are easily understood by reflecting on simple thought
experiments. The only reason why Einsteinian
relativity is somewhat less intuitive than its Galilean counterpart
is that the thought experiments on which it is based can be easily
referred to everyday experience
--- such as the original discussion by Galileo of moving
ships~\cite{galileo} --- while the thought experiments of relativity
require objects that move close to the speed of light. Indeed, the
idea that relativity would be part of the common-sense everyday
experience of living beings moving close to the speed of light has
been exploited in the classic popular presentation by
G.~Gamow~\cite{gamow}. 

Quantum mechanics is different. On the one hand, there
is overwhelming evidence that quantum mechanics provides the
underlying grammar for all fundamental physical laws, with no
evidence of deviations from it, either at a fundamental level (i.e.,
when shorter distances are tested) or at a macroscopic level (i.e.,
when the quantum behavior of systems consisting of an ever larger
number of degrees of freedom is tested). On the other hand, the
standard formulation of quantum mechanics requires introducing, in the
process of measurement, the interaction of the quantum system with a
classical measuring apparatus. 
Most physicist would
argue that this is surely an approximation, in that quantum mechanics
applies at all scales, and there is no such thing as a classical
realm. Certainly, as
mentioned, there is not a shred of experimental evidence of
macroscopic 
deviations
from quantum behavior which would allow for a separation between
quantum and classical worlds.  

But then, if the measurement process
as usually formulated is an approximation, it is an approximation to what?
There seems to be no full consensus on the
exact nature of the quantum to classical transition and therefore,
while most would agree that measurement  is driven by decoherence caused by
entanglement with a macroscopic measuring apparatus (see e.g. Sect.~20.3
of~\cite{schwabl} for a standard textbook presentation), there seems to
remain some fundamental underlying lack of understanding. The laws of quantum
mechanics are fundamentally strange, because an interpretation is
required in order to explain what they really mean. Indeed, there
exist numerous ``interpretations'' of quantum mechanics (see
e.g. 3.7 of~\cite{weinbergqm}): effectively, different interpretations
of quantum mechanics amount to an explanation of what the standard
set of quantum-mechanical rules are supposed to approximate --- with
the Copenhagen interpretation consisting of accepting the rules at
face value. 

Yet, in all contexts in which quantum mechanics is being used, the way
it should be used is absolutely clear. To the best of our knowledge,
there is not a single case in which the prediction of quantum
mechanics are ambiguous, even less a case in which an interpretation
is required in order to resolve the ambiguity. It is thus natural to
turn to the everyday use of quantum mechanics as a way to resolve its
strangeness: after all, if we know what the theory means in all
possible contexts, what more is there to say?

In the sequel, I will briefly discuss some aspects of quantum
mechanics in light of their use in the context of high-energy physics,
i.e. the physics of interactions at the shortest distance scales which
are probed in experiments such as those which are being performed at
the Large Hadron Collider of CERN~\cite{lhc}. I will show how many of these
aspects are taken for granted when using theoretical calculations to
obtain predictions which may be tested against experiment, and how
this makes many aspects of quantum mechanics look rather less paradoxical
than they might appear at first. We will suggest that this set of
ideas might perhaps lead to a full-fledged ``interpretation'' of
quantum mechanics, which we dub the ``Collider Interpretation of
Quantum Mechanics'':
though, in actual fact, what we are really talking about
is, somewhat more modestly,
the way of looking at quantum mechanics which is suggested by its use
in the context of high-energy physics. It will be left to the reader
to decide whether at present, or perhas at some later stage,
this deserves the noble name of ``interpretation''.
It should be clear from the onset, however, that we will not be
presenting any new results, and merely collecting some ideas that upon
reflection are
obvious: certainly obvious to any high-energy physicist. Also, many (perhaps
even all)  ideas discussed here are by no means unique to
high-energy physics, and could perhaps be presented in different contexts,
such as for instance nuclear physics: high-energy physics is meant to
provide a convenient, consistent conceptual framework.
 
\section{Randomness and collapse of the wave function}
\label{sec:random}

The basic quantities which are computed and compared to 
 experimental results in high-energy physics
are scattering cross-sections (and decay
rates). A fundamental feature of this 
comparison of theory with data is that it involves accepting the
fundamentally random nature of physical events. 

Take the simplest
scattering process, elastic electron-positron scattering (Bhabha
scattering), as it was done for instance 
at the LEP collider of
CERN~\cite{lep} (where it was, among other things, 
used to measure the luminosity
of the machine). Quantum electrodynamics, i.e. quantum mechanics applied to the
electromagnetic interaction, and its extension to also include the
weak interaction (the electroweak standard model) provide us with a prediction --- a very
precise prediction indeed --- for the
angular distribution of the outgoing electron-positron pair. 
cin each individual scattering event,
the outgoing particles exit in one particular direction. Their energy
is fixed by conservation of the incoming particle energy, and their
momenta must be equal and opposite by conservation of the total
momentum. But the exit angle is not fixed.  Quantum mechanics tells
us that if the experiment is repeated many times, we can predict on
average the
angular distribution of the outgoing particles (to astounding
accuracy), but not their exit angle in each
individual instance. This angular distribution is computed as a
function of the initial conditions, i.e. the momenta of the pair of
incoming, colliding particles, and it is fully determined by them.  
Unlike in a
classical scattering event, where uncertainty in the final state is
only related to imperfect knowledge of the initial conditions, or,
possibly, of the structure of colliding objects, this randomness has
to do with the fundamental nature of reality. Even when observing the
scattering of electrons (which are
pointlike object with no structure) with fully determined initial
momenta the exit direction cannot be predicted.

Once the particles do exit in one specific direction, we can take this
as initial condition for a subsequent quantum mechanical event. For
example, if the colliding particles were muons, instead of electrons
(muon colliders have not been built yet, but are actively
studied~\cite{mufact}) one might compute the chance that either of the
two outgoing muons would then decay. Quantum mechanics then provides
us with a determination of the probability that each muon may decay at any
given time, and the angular distribution and energy of its decays
products, as a function of the initial muon momentum, and indeed such
bread-and-butter calculations are routinely performed and used for
example in the calibration of particle detectors.

Summarizing, individual high-energy interactions are random: what the
laws of physics allow us to do is predict their behavior on average
as a function of the initial conditions. The outcome (the final state
of the interaction) can then be taken as a new initial condition.

It should be clear that this simple situation is at the root at the 
so-called ``collapse of
the wave function''. The wave function of the initial colliding
particle pair is the initial condition: it provides us with the
information on the state of the system\footnote{This information
might actually be incomplete; use of a density matrix formalism
allows a treatment this more general case, but this is besides our point
now}. 
Its knowledge only
allows us to compute probabilities of outcomes, which are then
 observables. After the observation of an actual outcome, our
 information on the state of the system (which we may use as initial
 condition for a subsequent measurement) changes. But the information
 on the state of the system is encoded in the wave function, so the
 wave function must change when we observe an actual individual
 outcome. After the measurement, the wave function no longer gives us
 a probability: rather, it is in the state which corresponds to the
 observed outcome. We can then use this new wave function to compute
 probabilities of subsequent events.

It is thus recognized that if one accepts that individual events are
fundamentally random, and their outcome can only be predicted on
average, then it follows that the information on the state of the
system changes discontinuously when an individual outcome is
observed.  Often  in popular
presentations this discontinuous change is described with statements
such as ``the observer perturbs the system''. But this 
seems at best to miss the main point, and in the worse
case rather misleading: when
observing the
outcome of an experiment (``performing a measurement'')
what   does  change, and discontinuously at that,  
is the observer's information 
on the state of the system.\footnote{The somewhat quaint description of the measurement as a
  perturbation of the system can be traced to Heisenberg (see
  e.g.~\cite{heisenberg}).}

Acceptance of this simple fact --- the fundamental randomness of
individual events --- helps in dispelling some of
the mystery that
sometimes appears to shroud many of the basic
tenets of quantum mechanics. For example, the postulate of quantum
mechanics that states that an observable is associated to a Hermitean
operator follows from the fact that the outcome of an experiment
(i.e. the outcome of a measurement) is not unique. In our example, 
we assign a different state of the system $|i\rangle$ to each
direction of the electron-positron pair. It is perhaps worth noticing
that in any realistic situation the set of such states is
always discrete. For example, a detector can measure the particle momenta
only with a certain resolution, so the outgoing angle can only take a
discrete set of values. Indeed, experimental
results for angular distributions are presented in the form of
histograms, in which the value of the momentum is provided in discrete
bins.\footnote{Clearly, this is also due to the fact that the number of
individual events on which each experimental measurement is based is
finite, but even in the limit of a very large number of events the
spacing of measured momentum values can never be finer than the detector
resolution.} So, after detecting the outgoing pair, the system is
found in one of the states $|i\rangle$, which acre each characterized
by a different
value of some observable, say an angle $\theta_i$. The system is to
be found with probability $p_i$ in state  $|i\rangle$. 

The observed mean value
of the angle is
\begin{equation}\label{eq:mean}
\langle \hat \theta\rangle= \sum_i \theta_i p_i 
\end{equation}
We can rewrite this by defining an operator $\hat \theta$, as
the matrix in the space spanned by states $|i\rangle$ which has values
$\theta_i$ on the diagonal. This definition is natural in that the mean value of
$\hat\theta$ is then found by tracing the product of this matrix with a density
matrix $\rho$, defined as the matrix which has the probabilities $p_i$
of outcomes in the diagonal:
\begin{equation}\label{eq:trace}
\langle \hat \theta\rangle= {\rm Tr} \hat \theta \rho.
\end{equation}

The Hermitean nature of $\hat \theta$ (and $\rho$) then simply follows
from the assumptions that the outcomes of the experiment are distinct
(i.e. orthogonal), and that they are each characterized by a value (not
necessarily distinct!) of a real number (the eigenvalues).
It should be clear that in this line of argument only the simple
principle of the statistical nature of experimental results has been
used: the observable is an operator because after the measurement there
is a distribution of distinct outcomes. This does still not fully
derive the Born rule - the fact that the state of the system can be
viewed as the superposition
\begin{equation}\label{eq:sup}
|\psi\rangle=\sum_i c_i |i\rangle
\end{equation}
such that $p_i=|c_i|^2$, though it is clear that Eq.~(\ref{eq:trace})
goes a long way in this direction.

We have thus understood the meaning of two basic axioms of quantum
mechanics --- the measurement and collapse of the wave function, and
observables as Hermitean operators --- and have a strong hint on the
nature of another one --- the Born rule. Many consequences which are
derived from them also appear perhaps less surprising and unusual, once
seen as statements about fundamentally random physical events of which
only the probability distribution can be predicted.

While many examples could be given, we consider here only one of the
simplest if not the simplest, namely  the
uncertainty principle. If fundamental physical events are random, then
their distribution is characterized not only by a mean value
Eq.~(\ref{eq:mean}), but also by a standard deviation
\begin{equation}\label{eq:sigma}
\Delta^2\hat \theta\equiv \sigma^2_\theta= \langle \hat
\theta^2\rangle -\langle\hat\theta\rangle^2 .
\end{equation}

The uncertainty principle is then merely a statement about the mutual
size of standard deviations of measurements of distinct observables for a
system in a given state. Needless to say, the standard deviation is a
property of the set of repetitions of the experiment, not of any
individual instance.
 It may be surprising that some
pairs of observables cannot be simultaneously sharp (because the
product of their standard deviations is bounded from below) in some
states, or perhaps even in all states. But 
it is clear that, again, this is not a
statement about what the observer does to the system, rather it is a
statement about an ensemble of repetitions of an experiment performed
on many identically prepared system.\footnote{This means that the uncertainty
principle relates the uncertainty of pairs of observables before a
measurement. The value of the uncertainty of an observable {\it after}
the measurement of another observable which does not commute witgh it
is a separate issue, and has in fact been studied recently~\cite{distler}}

\section{Superposition and scale separation}
\label{sec:super}

The Bhabha scattering process discussed in Sect.~\ref{sec:random} can
proceed through several ``channels'', as the jargon goes. Namely, the
incoming pair can produce either a photon or a $Z$, each of which can
then go into the given final state\footnote{More  intermediate states
  are possible in  higher orders of perturbation
theory.}. 
In fact, at LEP1 (the first phase
of running of LEP) the energy of the collision was tuned in such a way
that real $Z$s were being produced --- the energy of the collision was
tuned to be equal to the mass of the particle, so one could actually
view the $Z$Z channel as consisting of production of a $Z$ particle
followed by its subsequent decay. Yet, of course, the probability
for producing the final electron-positron pair is not obtained by
combining the probabilities  of going through the $Z$ and photon final
states, rather, by combining the amplitudes.

It might be interesting
to ask to which extent this postulate of quantum mechanics could
be derived starting with the argument of the previous section, based
on fundamental underlying randomness, and imposing further consistency
conditions\footnote{On top of these, the only remaining postulate of quantum
  mechanics is time evolution, i.e. the Schr\"odinger equation. 
This can be derived from the requirement that quantum mechanics
preserves the canonical structure, i.e. that time translations are
generated by the Hamiltonian, or, equivalently, that time translation
invariance implies energy conservation.}. Here however we will not
pursue this line of argument, and instead ask whether this --- the fact that
amplitudes, and not probabilities, are combined --- is problematic,
and if not, why not.

The fact that events are random does not seem paradoxical or
contradictory per se: one can simply conceive factual reality as a
sequence of events of which only the probability is determined a
priori. The future can be viewed as an infinitely branching tree, of
which only one branch is realized. The repeatability of situations
(initial conditions) makes predictivity possible, on a statistical if
not on a deterministic basis. 

However, the fact that amplitudes,
rather than probabilities, should be combined, does seem to entail  some
complications, in that it prevents the branching tree to be
observed with excessive resolution.  For example, in the classic
double slit experiment we cannot view the trajectory that leads to the
detection of a particle in a specific point of the screen as a
sequence of random events, the first of which leads the particle  
randomly (though with fixed probability) from the
source to one of the slits,
times a second event which leads it randomly (though with fixed
probability) from the slit to the screen: this would correspond to
composing probabilities, and it would not lead to the observed
interference pattern. 

Nevertheless, this can be done to some extent: one may define a
``consistent history''~\cite{griffiths}, i.e. a set of branchings such
that indeed the history of a given quantum system can be
viewed as a sequence of intermediate random events. Of course, this is
possible to the extent that the different branchings do not interfere
with each other. In other words, provided the branching tree is sufficiently
``coarse grained'' one can view the evolution of the system as a
sequence of random events, but if it is excessively fine grained, this
becomes impossible
because then alternative histories interfere, so there is no sense in
which only one of them could have been actually realized: if the
paths through the two slits interfere, then there is no sense in which
the particle could have actually passed through either.\footnote{However,
one may define a generalized
probability over the quantum evolution of a system which satisfies all
criteria of standard probability, but not positivity. Whenever the set
of probabilities of all individual branches do satisfy positivity,
then the histories are sufficiently coarse grained, i.e., the question
of which sequence of random event has led from the initial to the
final condition does have a meaning --- it is a ``settlable bet''~\cite{hartle}.}

Be that as it may, one might wonder why in practice in the application
to any high-energy physics experiment these issues never arise. One
never has to ask which is the random event: in the example above, it
is the elementary electron-positron scattering. One never has to ask what is
the initial condition: it is provided by the wave function of the
initial electron-positron pairs, which at a collider are momentum
eigenstates. And there is likewise no doubt about what is the
measurement: it is the observation of the outgoing electron and
positron in the detector, which, because the detector is designed to
determine the values of the energy and momentum of the outgoing
particles, leads to an initial condition for subsequent events --- a
final state wave function --- which also corresponds to momentum eigenstates.

The answer is clear upon a minutes reflection: scale separation is the
reason why one never has to deal with the issue of what is
the random event and what is the measurement, or, in the language of
the consistent histories, the issue of what are the projection
operators and what is the amount of coarse-graining which define the
consistent history. In our example, the scale which defines the
elementary interaction is that of the electroweak interaction: in a
scattering event at LEP it would be of the order of 100~GeV,
corresponding\footnote{With $\hbar c=1$, so $197$~MeV$\approx10^{13}$~cm$^{-1}$ (natural units).} to a length scale of about
$10^{-16}~cm^{-1}$. The initial states are not really momentum
eigenstates: they are wave packets, and in fact the standard
expression for the scattering cross-section involves a factor to
account for the flux of incoming particle derived (as explained
in any good textbook~\cite{peskin}) assuming the incoming particles to
be given by wave packets whose momentum uncertainty is small in
comparison to the momentum scale of
interaction, so their wave function is spread over
much larger distances (for example, 100~MeV, corresponding to
$10^{-14}~cm^{-1}$). The incoming particles are typically
prepared in a bunch, spread over much larger distances scales, and
with approximately constant particle density (a  proton bunch
at the LHC has a size of abou1 $10\mu m=10^{-3}~cm$). Likewise, the final
state is observed at distances which are macroscopically large in
comparison to the distance of the interaction: even the closest
detector, such as the pixel  detectors of LHC experiments observe
particles at distances of about $10 \mu m=10^{-3}~cm$ from the
interaction point. They reveal states whose momentum uncertainty is
again much smaller than that of the interaction which is being
studied, and which are seen by reconstructing tracks whose width
(i.e., position uncertainty) is typical  of distance scales of the
solid-state electronic devices which are used to detect them,
i.e. again a few $\mu m$ at most.

Therefore, the problem does not arise because the characteristic
distance scale of the
incoming and outgoing states, and that of the interaction, are
separated by many orders of magnitude. This, in turns, reflect the
setup of the devices which are used to prepare the incoming particles
and to detect the outgoing ones. The amount of coarse-graining is thus
fixed by the natural scales of the problem. 

Whereas we have illustrated this situation in a particular example,
it is in fact generic, and it provides the language in which
high-energy physics experiments are described and experiments are
designed, both from a technical, and a more informal point of view.
Indeed, the fact that ``in'' and ``out'' states are prepared at scales
which are so well separated from the interaction that they can be
described by a free theory is part of the basic underlying formalism
of quantum field theory~\cite{weinberg}. That ``in'' states are
momentum eigenstates from the point of view of the quantum-mechanical
computation, but really wave packet from the point of view of the
determination of a macroscopic cross-section is part of the way the
cross-section is computed: as already mentioned, the flux factor,
which is an integral part of the cross-section, is obtained using this
assumption in a crucial way. That the ``out'' states are momentum
eigenstates are measured by reconstruction of a `''particle
trajectory'' is an accepted fact which underlies experimental
design. Here particle trajectory is put in quotes because of course a
quantum-mechanical momentum eigenstates has completely uncertain
position and thus no trajectory. But in realistic experiments, as in
the above example, one deals with particles whose momenta are
measured with an uncertainty of, say tens or hundreds of MeV, by
measuring tracks whose width, and even less whose lengths, never go
below the $\mu m$ scale, at least six or seven orders of magnitude
below the limit from the uncertainty principle. Hence, one deals with
wave-packets, which may be treated as momentum eigenstates for all
practical purposes.

Hence, the  separation of scales between preparation, interaction, and
detection, guarantees first, that one may view the fundamental
interactions of high-energy physics as fundamentally random events
which are preceded and followed by well-define measurement processes
with no ambiguity, and second, that this measurement process, even
though fully quantum mechanical in nature (because the properties of a
single particle are measure, such as a single electron, and this is
surely a fully quantum-mechanical system) can be described in
semi-classical terms, as when stating that a particle's momentum is
determined by looking at the curvature of its trajectory when it is
subject to a magnetic field. 

Whereas we have only given the simplest example, it turns out that
this separation is at work even when this semi-classical language is
used to describe purely quantum-mechanical phenomena. For example,
neutrino oscillations~\cite{ridolfi} 
involve the time-evolution of states which are
the superposition of two different energy eigenstates, so that the
superposition coefficients depend on time. The standard
language used to describe them treats the eigenstates as particles
that propagate in space as a function of time, though, again, 
strictly speaking an energy-momentum eigenstate would have fully
uncertain position: so, in principle, only a description in terms of
wave packets would be correct~\cite{giunti}, but it turns out that
in practice the standard ``semiclassical'' description is fully
adequate~\cite{lipkin} because the scale of the interaction used to
prepare the incoming neutrino beam and to reveal the outgoing neutrino
states is characterized by enormously greater length scales in
comparison to that which drives the neutrino oscillation itself.

In summary, the language of ``particle flux'', ``interaction'' and
``particle detection'' separates the act of measurement with that of
quantum-mechanical evolution, without having to loose the
quantum-mechanical  nature of the incoming and outgoing states, and
even without fully erasing the quantum-mechanical nature of the act of
measurement itself.

\section{Multiple descriptions and preferred bases}
\label{sec:pointer}
We have seen that our basic example of a simple quantum scattering
process can be viewed as a random event which connects an initial
state in which the two incoming particles are in momentum eigenstates
to a final state in which a measurement also reveals them in momentum
eigenstates. It is clear, however, that the choice of basis states for
the description of the fundamental random event is not unique, and we
could have made the choice of describing the process, for instance,  in terms of
position eigenstates. 

There are well known examples in which the same quantum evolution
process may be viewed in different but completely equivalent
bases (for instance, for spin systems in which different spin
components are diagonalized~\cite{griffiths}). 
In each base it is possible to provide  a
sufficiently coarse-grained description, that the full evolution from
initial to final state is consistently expressed as a sequence of
random events --- so that at each intermediate state it is possible to
say which of the alternative histories is actually being followed by
the system. Equivalently, it is possible a posteriori to settle
the bet of which of the paths of the branching history tree the system
is following in a particular sequence of events. Yet, the descriptions
might be given in terms of bases of eigenstates of 
incompatible (i.e. non-commuting) operators, in such a way that
different inequivalent descriptions of the same history are
possible. This means that it is not possible to state that at some
intermediate time the system was objectively in one state, because
in, say, a pair of different, equivalent descriptions (based on an arbitrary basis
choice) the system was respectively in one or another state which
belong to incompatible bases. This means that one cannot say that, in
that history, the random event of the state being in one particular
eigenstate of either of these two bases ``actually happened''. But the
requirement that ``something happens'' has been
advocated~\cite{multiverse}, quite reasonably,  as a
requirement for a satisfactory formulation of quantum mechanics.

Once again, one may wonder why this issue never arises in
high-energy physics settings. In our example, it is clear that the
incoming and outgoing particles are in momentum eigenstates (well,
approximate eigenstates, really, as discussed in the previous
section), and that the transitions induced by the fundamental
interaction, which the Feynman diagram formalism treats using
standard time-dependent perturbation series and the Dyson series, are
transitions between different momentum eigenstates induced by
multiple insertions of the Hamiltonian. 

Specifically, one might wonder whether an equivalent 
description in terms of, say,
position eigenstates was instead possible --- perhaps with a slightly
different experimental setup such that position, instead of momentum
eigenstates are preferentially detected. If this were the case, one could say
that the very same elementary interaction has actually led to a
transition between position eigenstates instead, so that our naive
conviction that in each  scattering event a random transition between
different momentum eigenstates has happened is actually fictitious.

A minutes' reflection reveals that this is not the case, for a reason
which has to do with the fundamental structure of quantum field
theory, i.e. the quantum mechanics of space-time systems with infinite
degrees of freedom. Namely, in the nonrelativistic quantum
mechanics of one or many particles in space the coordinate and
momentum representations of the wave function are entirely equivalent:
we can expand the wave function over a set of position eigenstates, or
momentum eigenstates, and none is preferred over the other. Not so in
quantum field theory. Indeed, in (free) field theory the Hamiltonian
is diagonalized in the basis of momentum eigenstates: the Hamiltonian
in position space describes a system of coupled harmonic oscillators,
each localized at a point in space, which are diagonalized by normal
coordinates which are the momentum states. This is a consequence of
the structure of the free-field kinetic term as the operator with the
smallest number of spacetime derivatives, i.e., ultimately, it follows
from the fact that elementary excitations of space and time can always
described at the shortest distances as small oscillations about an
equilibrium state. 

The momentum degrees of freedom are supplemented 
by spin if one also considers the behavior of the fundamental
excitations upon spacetime rotations. In the high-energy limit 
all excitations are effectively
massless, and massless excitations are helicity eigenstates, i.e. they
have spin parallel to the momentum axis. Hence, even in the case of
spin degrees of freedom there is a preferred basis: than in which spin
is quantized along the direction of momentum. This is the only basis
which admits a well-defined short-distance limit. But the existence of
a short-distance limit is a basic requirement for any fundamental
theory, i.e. one that is meaningful at all length
scales.\footnote{Clearly, the same line of argument can be pursued
  when dealing with internal symmetries, where the scale separation
  of different interactions allows one to measure the degrees of
  freedom of one interaction --- such as, say, the isospin of
  strongly-interacting neutrons --- using decay through another
  interaction --- such as neutron $\beta$ mediated by the weak
  interaction. See below for another example in neutral kaon decay.}

We must conclude that the basis of momentum eigenstates, supplemented
by spin quantized in the direction of momentum, 
is selected by the fundamental interaction, for reasons which have to
do with the structure of space and time. Indeed, elementary particles
are eigenstates of mass and spin. This follows from the requirement
that elementary particles be eigenstates of the Casimir invariants of
the Poincar\'e group, which is the invariance group of spacetime.

It is interesting to observe that scale separation, as discussed in
the previous section, guarantees that the relevant momentum
eigenstates are always uniquely defined. For example, neutral kaons
are produced as mass eigenstates. However, the mass eigenstates are
not also eigenstates of the Hamiltonian which leads to their decay. As
a consequence of this, neutral kaons oscillate due to quantum
evolution, but the fact that the scale of their masses (hundreds of
MeV) 
is distinct
from the scale of the weak interactions (hundreds of GeV) guarantees
that the kaon decay, which selects the weak eigenstates, is well
separated from the propagation of the mass eigenstates. Indeed,
the two different weak eigenstates are distinguished by the length of
their trajectories (``long'' and ``short'') --- here one sees again at
work the semiclassical picture (as in the case of neutrino
oscillations) whereby one may talk of trajectory of a
quantum state, meaning that one is talking about a wave packet which
can be taken to be a momentum eigenstate from the point of view of the
fundamental underlying quantum interaction, because the momentum
uncertainty is much smaller than the position uncertainty.

Admittedly, the argument given here is by no means a general
proof. However, to the best of our knowledge,  the textbook
situation~\cite{griffiths}  in
which multiple incompatible descriptions of the same quantum history 
are possible has never arisen in the context of any realistic
high-energy physics experiment. Our line of argument suggests that
this is not accidental, but rather that it follows from deep properties
of quantum field theory which in turn reflect the way spacetime
symmetries are realized in the theories of fundamental interactions.

\section{An interpretation of quantum mechanics?}

In this essay, we have addressed the problem of the interpretation of
quantum mechanics by asking why an interpretation does not seem to be
necessary at all when applying quantum mechanics to high-energy 
physics,  i.e., when using quantum mechanics to study the theory of
``fundamental'' interactions, namely, those which describe physics at
the shortest distances. We have argued that several tenets
(``postulates'') of quantum mechanics follow directly from recognizing
the fundamental randomness of basic (``interaction'') events. These
include the collapse of the wave function and the identification of
observables with Hermitean operators. Furthermore, we have argued that
the description of quantum mechanical evolution in terms of random
events which connect distinguishable states (``consistent histories'')
is always possible in this context because of the natural scale
separation of fundamental interaction events which in turn follows from
the structure of the interaction itself. Finally, we have argued that
this consistent history description is unique (so that one may view
the events which form it as having ``actually happened'') because the
basic symmetries of spacetime select momentum (over position) and
helicity (over other components of spin) as preferred bases. 

The fact that these considerations apply to the theory of fundamental
interactions suggests that their interest is more than
anectodal. Indeed, standard Wilsonian renormalization group
argument~\cite{weinberg} suggests that, indeed, the physics at the
shortes distance scales is ``fundamental'' in the sense that
lower-energy physics can be derived from it in the form of an
effective theory, by integrating out degrees of freedom. If one can
put on a firm footing an argument which shows that no interpretive
issues arise when applying quantum mechanics at these scales, then
this suggests that such interpretive issues might be an artefact of
the low-energy effective description.

Of course, this would require first, turning the simple arguments
presented here into more formal proofs, and second, filling in the
numerous gaps (e.g. by showing how also the Born rule is necessary,
and so forth). We call the completion of this program the
``High-Energy Interpretation of Quantum Mechanics'' --- effectively, a
proof that no interpretation of quantum mechanics is really needed.

\end{EJarticle}

\end{document}